\documentclass[final]{siamltex}

\usepackage[cmex10]{amsmath}
\usepackage{amssymb,amsfonts}
\usepackage{mathdots}
\usepackage{moreverb}

\title{Derivation of fast DCT algorithms using algebraic technique based on Galois theory\footnotemark[2]}

\author{Maxim Vashkevich\footnotemark[3] \and Alexander Petrovsky\footnotemark[3]}

\begin{document}

\maketitle

\renewcommand{\thefootnote}{\fnsymbol{footnote}}

\footnotetext[2]{This work was supported by the Belarusian Fundamental Research Fund (F11MS-037).}
\footnotetext[3]{Dept. of Computer Engineering, Belarusian State University of Informatics and Radioelectronics, Minsk, Belarus, 220013, vashkevich@bsuir.by.}

\renewcommand{\thefootnote}{\arabic{footnote}}

\begin{abstract}
The paper presents an algebraic technique for derivation of fast discrete cosine transform (DCT) algorithms. The technique is based on the algebraic signal processing theory (ASP). In ASP a DCT associates  with a polynomial algebra $\mathcal{A}_\mathbb{C}=\mathbb{C}[x]/p(x)$. A fast algorithm is obtained as a stepwise decomposition of $\mathcal{A}_\mathbb{C}$. In order to reveal the connection between derivation of fast DCT algorithms and Galois theory we define $\mathcal{A}$ over the field of rational numbers $\mathbb{Q}$ instead of complex $\mathbb{C}$. The decomposition of $\mathcal{A}_\mathbb{Q}$ requires the extension of the base field $\mathbb{Q}$ to splitting field $\mathbb{E}$ of polynomial $p(x)$. Galois theory is used to find intermediate subfields $\mathbb{L}_i$ in which polynomial  $p(x)$ is factored. Based on this factorization fast DCT algorithm is derived.
\end{abstract}



\pagestyle{myheadings}
\thispagestyle{plain}
\markboth{M.~VASHKEVICH AND A.~PETROVSKY}{FAST DCT ALGORITHMS BASED ON GALOIS THEORY}

\section{Introduction}
The DCT~\cite{Strang/99} is the most widely used transforms in video and image processing, especially in coding application~\cite{Rao2006}. Different algebraic structures of DCT's have been investigated  by several authors~\cite{Steidl1991,Feig1997}{} in order to obtain bounds on the arithmetic complexity of their computations~\cite{Feig1992} and derive fast algorithms~\cite{Chich1997}. The most general and comprehensive theory that explains the relation between DCT and algebraic structures has been presented in~\cite{Pusch3}. Subsequently this theory has been called {\it algebraic signal processing theory} (ASP)~\cite{Pusch8d}. A characteristic feature of ASP is that underlying {\it signal model} is considered instead of a transformation matrix. The notion of  signal model implies a triple $(\mathcal{A,M},\Phi)$, where $\mathcal{A}_\mathbb{C}=\mathbb{C}[x]/p(x)$ is an algebra of filters, $\mathcal{M}$ is an $\mathcal{A}$-module of signals (or signal space) and $\Phi$ is a generalized concept of $z$-transform. Here and below we use subscript to indicate the base field of the polynomial algebra where it is necessary.

The correspondence between a polynomial algebra and DCT is a key point of ASP. A fast transform algorithm is obtained as a stepwise decomposition of $\mathcal{A}_\mathbb{C}$, that requires a stepwise factorization of $p(x)$~\cite{Sandryhaila:2011:ASP}. We propose to define $\mathcal{A}$ over the field of rational numbers $\mathbb{Q}$. Note that except trivial cases $p(x)$ cannot be factored over $\mathbb{Q}$ and therefore we need to extend the base field to splitting field $\mathbb{E}$ of $p(x)$. 

The stepwise factorization of $p(x)$ and the extension of based field $\mathbb{Q}$ are carried out using Galois theory. The theory allows us to find a set of intermediate subfields $\mathbb{L}_i$ between $\mathbb{Q}$ and $\mathbb{E}$.  In each $\mathbb{L}_i$ the polynomial  $p(x)$ is stepwise factored. The factorization results in a fast DCT algorithm. 

As an application of the proposed technique we derive a fast recursive algorithm for $\mathrm{DCT}\text{-}4_{n}$, where $n=2^k$.

\section{Algebraic background}
An algebra is a vector space $\mathcal{A}$ over some base field $\mathbb{F}$ in which multiplication of elements is defined and satisfies distributivity law. For instance complex numbers, quaternions and $\mathbb{Q}[x]$ (set of polynomials with rational coefficients) are algebras.

The basic notion of ASP is {\it polynomial algebra} $\mathcal{A}_\mathbb{C}=\mathbb{C}[x]/p(x)$. If $p(x)$ is polynomial of degree $n$ then $\mathbb{C}[x]/p(x)=\{ q(x) \mid \deg(q)<n\}$ is a set of polynomials of degree smaller then $n$ with addition and multiplication modulo $p(x)$.

Polynomial algebra $\mathbb{C}[x]/p(x)$ can be decomposed into a direct sum of irreducible subalgebras $\mathbb{C}[x]/(x-\alpha_k)$ using the {\it Chinese remainder theorem (CRT)}:
\begin{equation*}
\mathcal{F} \; \colon \;\; \mathbb{C}[x]/p(x)\rightarrow \bigoplus_{0\leq k < n} \mathbb{C}[x]/(x-\alpha_k).
\end{equation*} 	
under assumption that zeros $\alpha=(\alpha_0,\ldots,\alpha_{n-1})$ of $p(x)$ are pairwise distinct.

The operator $\mathcal{F}$ can be expressed in the following matrix form
\begin{equation}
\mathcal{F} = \mathcal{P}_{b,\alpha}=[p_\ell(\alpha_k)]_{0\leq k,\ell <n},
\label{eq:sig_mod_Fourier}
\end{equation}
where $b=(p_0,\ldots,p_{n-1})$ is a basis in $\mathbb{C}[x]/p(x)$. Eq.~(\ref{eq:sig_mod_Fourier}) assumes that in each  $\mathbb{C}[x]/(x-\alpha_k)$ the unit bases is set $(x^0)=(1)$.
$\mathcal{P}_{b,\alpha}$ is a {\it polynomial transform} for $\mathcal{A}_{\mathbb{C}}$ with basis~$b$~\cite{Pusch8d}. A {\it scaled polynomial transform} is obtained for a different basis $\beta_k$ in each $\mathbb{C}[x]/(x-\alpha_k)$:
\begin{equation}\label{eq3}
\mathcal{F} =\diag(1/\beta_1,\ldots,1/\beta_{n-1}) \cdot \mathcal{P}_{b,\alpha}.
\end{equation}

\subsection{Algebraic derivation of fast transform algorithm in ASP}\label{ASP_fw}
As stated above fast transform algorithm in ASP is derived as a stepwise decomposition of underlying algebra $\mathbb{C}[x]/p(x)$ into a direct sum of one-dimensional algebras. Let $p(x)=q(x)\cdot r(x)$, $k=\deg(q)$ and $m=\deg(r)$ then
\begin{eqnarray}
&& \mathbb{C}[x]/p(x) \notag \\
& \rightarrow &  \mathbb{C}[x]/q(x)\oplus \mathbb{C}[x]/r(x) \label{eq6}\\
& \rightarrow & \bigoplus\limits_{0\leq i<k}\mathbb{C}[x]/(x-\beta_i) \oplus \bigoplus\limits_{0\leq j<m} \mathbb{C}[x]/(x-\gamma_j) \label{eq7}\\
& \rightarrow & \bigoplus\limits_{0\leq i<n}\mathbb{C}[x]/(x-\alpha_i) \label{eq8}
\end{eqnarray}
where $\beta_i$ and $\gamma_j$ are the zeros of $q(x)$ and $r(x)$ correspondingly. If $c$ and $d$ are the bases of $\mathbb{C}[x]/q(x)$ and $\mathbb{C}[x]/r(x)$, respectively, then (\ref{eq6})-(\ref{eq8}) are expressed in the following matrix form~\cite{Pusch8d}:
\begin{equation}\label{th1}
\mathcal{P}_{b,\alpha} = P(\mathcal{P}_{c,\beta}\oplus \mathcal{P}_{d,\gamma})B,
\end{equation}
where $A\oplus B=[\begin{smallmatrix} A & \\ & B\end{smallmatrix}]$ denotes the direct sum of matrices.
The matrix $B$ maps the basis $b$ to concatenation of the bases $(c,d)$ and corresponds to (\ref{eq6}). Eq. (\ref{eq7}) uses the CRT to decompose $\mathbb{C}[x]/q(x)$ and $\mathbb{C}[x]/r(x)$. This step corresponds to the direct sum of matrices $\mathcal{P}_{c,\beta}$ and $\mathcal{P}_{d,\gamma}$. Finally permutation matrix $P$ maps the concatenation $(\beta,\gamma)$ to the ordered list of zeros $\alpha$ in (\ref{eq8}). Given that $B$ is sparse~(\ref{th1}) leads to a fast algorithm. 

\subsection{Polynomial algebra for $\mathrm{DCT}$-$4$}
Let us consider the polynomial algebra associated with widely used $\mathrm{DCT}\text{-}4_n$
\begin{equation}
\mathcal{A}_\mathbb{C} = \mathbb{C}[x]/2T_n(x),\quad b=(V_0,\ldots,V_{n-1}), \label{dct4_model_a}
\end{equation}
where 
$T$ and $V$ are Chebyshev polynomials of the first and third kind, respectively. Chebyshev polynomials have the following closed form expressions ($\cos\theta=x$)
\begin{equation*} \textstyle
T_n(x) = \cos(n\theta),\,\,\, V_n(x) = {\cos(n+\frac12)\theta}\Big/{\cos\frac12\theta}.
\end{equation*}
$\alpha_k=\cos(k+\frac12)\frac{\pi}{n}$, $0\leq k<n$ are zeros of $2T_n(x)$. Thus, according to (\ref{eq:sig_mod_Fourier}) polynomial transform for algebra (\ref{dct4_model_a})  is defined as
\begin{equation} \label{eq4}
\mathcal{P}_{\alpha,b} = [ V_\ell(\alpha_k)]_{0\leq k,\ell<n} = \left[  \frac{\textstyle \cos(k+\frac{1}{2})(\ell+\frac{1}{2})\frac{\pi}{n}} {\textstyle\cos(k+\frac{1}{2})\frac{\pi}{2n}} \right].
\end{equation}
In order to get the matrix of $\mathrm{DCT}$-$4_n$ (\ref{eq4}) is multiplied  from the left by scaling diagonal matrix
\begin{equation} \label{eq:sc_mtx}\textstyle
D^{(C4)}_n = \diag\nolimits_{0\leq k< n} \left({\cos(k+\frac12)\frac{\pi}{2n}}\right)
\end{equation}
that yields
\begin{equation} \label{eq5}\textstyle
\mathrm{DCT}\text{-}4_n =\left[\cos(k+\frac12)(\ell+\frac12)\frac{\pi}{n} \right]_{0\leq k,\ell<n}.
\end{equation}
Eq. (\ref{eq4})--(\ref{eq5}) show that $\mathrm{DCT}$-$4$ is a scaled polynomial transform of the form (\ref{eq3}) for the specified polynomial algebra~(\ref{dct4_model_a}). 

Polynomial transform corresponding to discrete trigonometric transform ($\mathrm{DTT}$) is denoted as $\overline{\mathrm{DTT}}$ (for instance $\overline{\mathrm{DCT}\text{-}4}_n$ stands for the matrix in (\ref{eq4})). 

\section{Changing base field of polynomial algebra}
Special feature of the polynomial algebras $\mathcal{A}_\mathbb{C} = \mathbb{C}[x]/p(x)$ associated with DCT is that all coefficients of polynomial $p(x)$ are integer~\cite{Pusch8d}. 
Thus the base field of $\mathcal{A}_\mathbb{C}$ can be changed to the field of rational numbers $\mathbb{Q}$: 
\begin{equation}
\mathcal{A}_\mathbb{Q} = \mathbb{Q}[x]/p(x). 
\end{equation}
In the general case complete factorization of $p(x)$ requires an extension field $\mathbb{Q}$.

\subsection{Field extensions and splitting field}
	A field $\mathbb{E}$ is an {\it extension field} of the field $\mathbb{F}$ if $\mathbb{F}\subseteq \mathbb{E}$. The extension field $[\mathbb{E}\colon \mathbb{F}]$ is a {\it splitting field} of the polynomial $p(x)$ over $\mathbb{F}$ if $\mathbb{F}\subseteq \mathbb{E}$, $p(x)$ splits over $\mathbb{E}$ and $\mathbb{E}$ is generated by the roots of $p(x)$. 

{\bf Example}. $\mathbb{Q}[\sqrt{2}]$ is the extension field of $p(x)=x^2-2$. The field $\mathbb{Q}[\sqrt{2}]$  is generated by the adjunction of the element $\sqrt{2}$ to $\mathbb{Q}$. 

One-to-one mapping from $\mathbb{E}$ onto $\mathbb{E}$ is referred to as {\it automorphism}. Let $\mathbb{E}$ be an extension field of $\mathbb{F}$, then  $\mathbb{F}$-automorphism of $\mathbb{E}$ is an automorphism $\phi$ of $\mathbb{E}$ such that $\phi(x)=x,\,\forall x\in \mathbb{F}$. The group of such automorphisms is denoted by $\mathrm{Aut}\,\mathbb{E}/\mathbb{F}$.

{\bf Example}. Let us define the function $f\colon\,\mathbb{Q}[\sqrt{2}]\rightarrow\mathbb{Q}[\sqrt{2}]$ such that 
\begin{equation}
f(a+b\sqrt{2})=a-b\sqrt{2}, \label{eq:f_auto}
\end{equation}
then $f$ is $\mathbb{Q}$-automorphism of the field $\mathbb{Q}[\sqrt{2}]$.

\subsection{Foundations of Galois theory}
Galois theory can be used for finding extensions of $\mathbb{Q}$ needed to factor $p(x)$. This extensions are defined by corresponding subgroups of Galois group of $p(x)$.

Let us consider a polynomial $p(x)\in \mathbb{F}[x]$ with distinct roots. If $\mathbb{E}$ is a splitting field of $p(x)$ then the group $\mathrm{Aut}\,\mathbb{E}/\mathbb{F}$ is called {\it Galois group} of $p(x)$ and denoted by $\mathrm{Gal}(\mathbb{E}/\mathbb{F})$ or $\mathrm{Gal}_{p}$.

{\bf Example}.
The splitting field of polynomial $p(x)=x^2-2$ is $\mathbb{Q}[\sqrt{2}]$. 
Galois group of $p(x)$ consists of two elements $\mathrm{Gal}(\mathbb{Q}[\sqrt{2}] / \mathbb{Q}])=\left\{ f,e \right\}$, where $f$ is given in (\ref{eq:f_auto}), $e$ is the identity element of the group $e(a+b\sqrt{2})=a+b\sqrt{2}$ and $f\cdot f=e$. Thus $\mathrm{Gal}(\mathbb{Q}[\sqrt{2}] / \mathbb{Q})$ is a cyclic group of order~2.

The main notion that is used below is {\it Galois correspondence} between intermediate fields of field extension $[\mathbb{E\colon F}]$ and subgroups of its Galois group. Each subgroup $H\in\mathrm{Gal}(\mathbb{E/F})$ corresponds to a subfield $\mathbb{L}\subset\mathbb{E}$, which is given by
\begin{equation*}
\mathbb{L}=\left\{ x\in\mathbb{E} \mid \phi(x)=x,\,\forall \phi\in H \right\}.
\end{equation*}
The fundamental theorem of Galois theory states that any tower of fields
\begin{equation}
\mathbb{F}=\mathbb{L}_0 \subset \mathbb{L}_1 \subset \cdots \subset \mathbb{L}_r = \mathbb{E} \label{eq:tower}
\end{equation}
corresponds to a normal series of subgroups of Galois group $\mathrm{Gal}(\mathbb{E/F})$ in reverse order
\begin{equation}
\mathrm{Gal}(\mathbb{E/F})=G_0\supset G_1 \supset \cdots \supset G_r=\{1\}.
\end{equation}
This bijection between subfields $\mathbb{L}_i$ and subgroups $G_i$ is called Galois correspondence.

\subsection{Using Galois correspondence for derivation of fast transform algorithm}

For an irreducible polynomial $p(x)\in\mathbb{F}[x]$ with a splitting field $\mathbb{E}$ the tower of fields (\ref{eq:tower}) can be obtained using Galois correspondence. $p(x)$ is decomposed into factors of a lower degree in each subsequent field $\mathbb{L}_i$ and finally represented as a product of linear factors $p(x)=\prod_{i=0}^{n-1}(x-\alpha_i)$ in $\mathbb{E}$. Considering (\ref{th1}) the process results in a fast algorithm.

\section{Derivation of fast $\mathrm{DCT}\text{-}4_{2^k}$ algorithm}
In this section foregoing theoretical notions are applied to the derivation of fast recursive $\mathrm{DCT}\text{-}4_{2^k}$ algorithm. Polynomial algebra associated with $\mathrm{DCT}\text{-}4_{2^k}$ is defined as
\begin{equation*}
\mathcal{A}_\mathbb{Q} = \mathbb{Q}[x]/2T_{2^k}(x).  
\end{equation*}
Let us define Galois group $\mathrm{Gal}_{2T_{2^k}}$ of $2T_{2^k}(x)$. 
$\mathrm{Gal}_{2T_{2^k}}$ is a cyclic groups of order $2^k$ ($\mathsf{Z}_{2^k}$):
\begin{equation}
\mathrm{Gal}_{2T_{2^k}} \cong \mathsf{Z}_{2^k}.
\label{Gal_T_2^k}
\end{equation}
The last equation leads us to the following normal series of subgroups of  $\mathrm{Gal}_{2T_{2^k}}$:
\begin{equation}
\mathrm{Gal}_{2T_{2^k}} \cong \mathsf{Z}_{2^k} \supset \mathsf{Z}_{2^{k-1}} \supset \cdots \supset \mathsf{Z}_2 \supset \{ 1 \},
\end{equation}
and corresponding tower of field 
\begin{equation} \textstyle
\mathbb{Q} \subset \mathbb{Q}\left[\sqrt{2}\right] \subset \cdots \subset \mathbb{Q}\left[\underbrace{\sqrt{2 + \sqrt{2 +  \ldots +\sqrt{2}}}}_{k}\right]. \label{eq:tower_T_2^k}
\end{equation}

Using (\ref{Gal_T_2^k})--(\ref{eq:tower_T_2^k}) the factorization of $2T_{2^k}(x)$ can be expressed as a recursive formula.
In order to get its general form let us consider the special case $k=2$.
The zeros of $2T_4(x)$ are $\alpha_\ell = \cos(\ell+\frac12)\frac{\pi}{4}$, $\ell=0,\ldots,3$: 
\begin{eqnarray*}
\alpha_0 &=  \cos\frac{\pi}{8}=-\alpha_3=-\cos\frac{7\pi}{8}&=\textstyle\frac12\sqrt{2+\sqrt{2}}, \\
\alpha_1 &=  \cos\frac{3\pi}{8}=-\alpha_2=-\cos\frac{5\pi}{8}&=\textstyle\frac12\sqrt{2-\sqrt{2}}.
\end{eqnarray*}
The splitting field of $2T_4(x)$ is $\mathbb{Q} \left[ \sqrt{2+\sqrt{2}}\right]$. 
In this case any element $\theta$ of the field is expressed as
\begin{equation*} \textstyle
\theta = u + v\sqrt{2} + w\sqrt{2-\sqrt{2}} + q\sqrt{2+\sqrt{2}},
\end{equation*}
where $u,v,w,q\in\mathbb{Q}$. 
The field is a four-dimensional vector space over $\mathbb{Q}$.

Group $\mathrm{Aut} \mathbb{Q}\left[ \sqrt{2+\sqrt{2}}\right]/\mathbb{Q}$ is the Galois group of $2T_4(x)$. Its elements are the following automorphisms:
\begin{align*}
\sigma_0(\theta) &=\textstyle u+v\sqrt{2}+w\sqrt{2-\sqrt{2}} + q\sqrt{2+\sqrt{2}}, \\
\sigma_1(\theta) &=\textstyle u-v\sqrt{2}-w\sqrt{2+\sqrt{2}} + q\sqrt{2-\sqrt{2}}, \\
\sigma_2(\theta) &=\textstyle u-v\sqrt{2}+w\sqrt{2+\sqrt{2}} - q\sqrt{2-\sqrt{2}}, \\ \sigma_3(\theta) &=\textstyle u+v\sqrt{2}-w\sqrt{2-\sqrt{2}} - q\sqrt{2+\sqrt{2}}.
\end{align*}

The Galois group $\mathrm{Gal}_{2T_4}$ is presented in table~\ref{tab:Gal_T_4}. 

\begin{table}[t]
\footnotesize
\renewcommand{\arraystretch}{1.3}
\caption{Galois group $\mathrm{Gal}_{2T_4}$} 
\label{tab:Gal_T_4}
\centering
\begin{tabular}{c|cccc}
$\cdot$ 	&	$\sigma_0$	&	$\sigma_1$	&	$\sigma_2$	&	$\sigma_3$ 	\\
\cline{1-5}
$\sigma_0$ 	&	$\sigma_0$	&	$\sigma_1$	&	$\sigma_2$	&	$\sigma_3$	\\
$\sigma_1$ 	&	$\sigma_1$	&	$\sigma_3$	&	$\sigma_0$	&	$\sigma_2$	\\
$\sigma_2$ 	&	$\sigma_2$	&	$\sigma_0$	&	$\sigma_3$	&	$\sigma_1$	\\
$\sigma_3$ 	&	$\sigma_3$	&	$\sigma_2$	&	$\sigma_1$	&	$\sigma_0$	
\end{tabular}
\end{table}

The table shows that $H\in \{\sigma_0,\sigma_3\}$ is a subgroup of $\mathrm{Gal}_{2T_4}$,  $\mathrm{Gal}_{2T_4}\cong\mathsf{Z}_4$ and $H\cong\mathsf{Z}_2$. The subgroup $H$ determines the intermediate subfield $\mathbb{Q}_H$ 
\begin{equation*} \textstyle
\mathbb{Q}_H =\{ \theta \in \mathbb{Q} \left[ \sqrt{2+\sqrt{2}}\right] \mid g(\theta)=\theta\;\; \forall g \in H \}.
\end{equation*}

It is obvious that $\mathbb{Q}_H\cong \mathbb{Q}[\sqrt{2}]$ and by Galois correspondence
\begin{equation*} \textstyle
\mathsf{Z}_4  \supset  \mathsf{Z}_2  \supset   \{ 1\} \Rightarrow 
\mathbb{Q} \subset  \mathbb{Q}\left[\sqrt{2}\right]  \subset \mathbb{Q}\left[\sqrt{2 + \sqrt{2}}\right].
\end{equation*}
The factorization of $2T_4$ is given below:
\begin{align*}
\textstyle \mathbb{Q}&\colon  2T_4(x), \notag\\
\textstyle \mathbb{Q}[\sqrt{2}]&\colon  \textstyle (2T_2(x)-\sqrt{2})(2T_2(x)+\sqrt{2}), \notag\\
\textstyle \mathbb{Q}{\scriptstyle \left[\sqrt{2 + \sqrt{2}}\right]}&  \colon  \textstyle (2T_1(x)+\sqrt{2+\sqrt{2}})(2T_1(x)-\sqrt{2+\sqrt{2}})  \notag \\
&\textstyle  \phantom{\colon}\,(2T_1(x)+\sqrt{2-\sqrt{2}})(2T_1(x)-\sqrt{2-\sqrt{2}}). \notag
\end{align*}

\subsection{Recursive factorization of $2T_{2^k}$}

From the special case it can be induced that $T_{2^k}$ is factored onto polynomials of the form $2T_n - 2\cos r\pi$ and the factorization process is expressed by the following recursive formula:
\begin{equation}
\textstyle 2T_{2^k}(x) -2\cos r\pi = \mathstrut\left(2T_{2^{k-1}}(x)-2\cos\frac{r\pi}{2}\right) 
\left(2T_{2^{k-1}}(x)-2\cos \pi(1-\frac{r}{2})\right), \label{eq:gen_fact}
\end{equation}
that can be proved using the closed form of $T_{2^k}$.

The formula defines factorization of $2T_{2^k}(x)$ when $r=1/2$. Because of $2\cos\frac{r\pi}{2} = \sqrt{2+2\cos r\pi}$ the left and right sides of (\ref{eq:gen_fact}) have different base fields: $\mathbb{Q}[2\cos r\pi]$ and $\mathbb{Q}[\sqrt{2+2\cos r\pi}]$ respectively. Equation (\ref{eq:gen_fact})  provides successive transition from $\mathbb{Q}$ to $\mathbb{E}$ in accordance with (\ref{eq:tower_T_2^k}).

\subsection{Recursive fast $\mathrm{DCT}$-$4_{2^k}$ algorithm}

In this section a fast recursive $\mathrm{DCT}\text{-}4_{2^k}$ algorithm is derived. The starting point is the polynomial algebra\footnote{Here $\mathbb{Q}_{\cos r\pi}$ is used as a short notation for field extension $\mathbb{Q}[\cos r\pi]$.} 
\[ \mathcal{A}_{\mathbb{Q}_{\cos r\pi}}=\mathbb{Q}_{\cos r\pi}[x]/(2T_{2^k}(x)-2\cos r\pi)\] 
with the basis $b=(V_0,\ldots,V_{2^k-1})$, that corresponds to skew $\mathrm{DCT}\text{-}4_n(r)$~\cite{Pusch8d}, where $0<r<1$. The conventional $\mathrm{DCT}\text{-}4_n$ is the special case of skew $\mathrm{DCT}\text{-}4_n(r)$ for $r=1/2$.

Using (\ref{eq:gen_fact}) the algebra $\mathcal{A}_{\mathbb{Q}_{\cos r\pi}}$ can be decomposed into the direct sum of subalgebras
\begin{eqnarray}
&& \mathbb{Q}_{\cos r\pi}[x]/(2T_{2^k}(x)-2\cos r\pi) \notag \\
& \rightarrow &  \textstyle \mathbb{Q}_{\cos\frac{r\pi}{2}}[x]/(2T_{2^{k-1}}(x)-2\cos\frac{r\pi}{2}) \oplus \notag\\
&& \textstyle \mathbb{Q}_{\cos\frac{r\pi}{2}}[x]/(2T_{2^{k-1}}(x)-2\cos \pi(1-\frac{r}{2})). \label{eq:my_step1} 
\end{eqnarray}
Considering ASP framework described in \S~\ref{ASP_fw} and choosing basis $c=d=(V_0,\ldots,$ $V_{2^{k-1}-1})$ in the subalgebras the following fast $\overline{\mathrm{DCT}\text{-}4}_{2^k}(r)$ algorithm is derived:
\begin{equation}\label{my_alg}
\textstyle 
\overline{\mathrm{DCT}\text{-}4}_{2^k}(r) =  P\cdot( \textstyle\overline{\mathrm{DCT}\text{-}4}_{2^{k-1}}(\frac{r}{2})  \oplus\textstyle\overline{\mathrm{DCT}\text{-}4}_{2^{k-1}}(1-\frac{r}{2}) )\cdot B^{(C4)}_{2^k},
\end{equation}
where $P$ is a permutation matrix of the form
\begin{equation*}
P= \left[
\begin{smallmatrix} 
	1	&		& 			&	&			&			&		\\
		&		&			& 	&	I_2	& 			&		\\
		&	I_2&			&	&	 		&			&		\\
		&		&			&	&	 		&\ddots	&		\\
		& 		&\ddots	&	&			&			&		\\
		&		&			&	&			&			&	I_2\\
		&		&			& 1&			&			&	
\end{smallmatrix}\right],
\end{equation*}
and $B^{(C4)}_{2^k}$ is the change of basis matrix from the basis $b$ to the concatenation  $(c,d)$. The elements $V_\ell\in b$ for $0\leq\ell<m=2^{k-1}$ are actually contained in $c$ and $d$. $V_{m+\ell}$ are expressed in the new basis
\begin{equation}
\begin{split}
V_{m+\ell} &\equiv\textstyle -  V_{m-\ell-1} +  \textstyle 2\cos\frac{r\pi}{2}V_\ell
		 \mod{2T_{m} - 2\cos\frac{r\pi}{2}}  
\\ \textstyle
V_{m+\ell} &\equiv\textstyle -  V_{m-\ell-1} - \textstyle 2\cos\frac{r\pi}{2}V_\ell 
		 \mod{2T_{m} - 2\cos \pi(1-\frac{r}{2})}.  \label{eq_new_V}
\end{split}
\end{equation}
The equation can be induced from $2T_m=V_m+V_{m-1}$ and recurrence for Chebyshev polynomial $V_{n} = 2xV_{n-1} - V_{n-2}$. 
Given (\ref{eq_new_V}) 

\begin{equation*}
B^{(C4)}_{2^k}
= \begin{bmatrix}
I_m	&	(aI_m-J_m)	\\
I_m	&	(-aI_m-J_m)
\end{bmatrix} 
=\begin{bmatrix}
I_m	&	I_m	\\
I_m	&	-I_m
\end{bmatrix}\cdot
\begin{bmatrix}
I_m	&	-J_m	\\
		&	aI_m
\end{bmatrix},
\end{equation*}
where $I_m$ and $J_m$ is identity and reverse identity $m\times m$ matrices and $a= 2\cos\frac{r\pi}{2}$.

Finally the fast algorithm of conventional $\mathrm{DCT}\text{-}4$ is obtained by multiplying  (\ref{my_alg}) from the left by matrix $D^{(C4)}_n$ defined in (\ref{eq:sc_mtx}). 
\section{Discussion}
The derived fast algorithm is a special case of the algorithm obtained in~\cite[see eq. (48)]{Pusch8d}. It should be noted that above mentioned algorithm is based on the decomposition property $T_{km}=T_k(T_m)$ of Chebyshev polynomial. We have shown that the same algorithm can be obtained in a different way by using the factorization (\ref{eq:gen_fact}). An important result is the establishment of connection between Galois theory and the structure of fast DCT algorithm.

Scaled version of obtained $n$-point fast DCT-4 algorithm (were $n=2^k$) requires $\frac{n}{2}\log_2{n}$ multiplications and $\frac{3n}{2}\log_2{n}$ additions. In order to scale the outputs $n$ extra  multiplications are needed. Note that in video and image coding applications the output of the DCT can usually be scaled since it will be followed by a quantizer that can take this scaling into account. 

A scaled version of 8-point fast DCT-2 that requires only 5 multiplication can be obtained using the proposed fast DCT-4 algorithm. Algorithm with the same arithmetic complexity is obtained in~\cite{Arai88}. However, using the proposed DCT-4 algorithm the method of derivation scaled versions of fast $\mathrm{DCT}\text{-}2_{2^k}$ algorithms with low multiplicative complexity is generalized (see the appendix for detail).

Another interesting application of proposed algorithm is development of error-free computational algorithms of fast DCT. In~\cite{Dimitrov05} a method of error-free computation of fast DCT based on Arai algorithm with algebraic integer (AI) encoding is presented. The proposed algebraic technique allows to express the quantity in each node of graph of DCT algorithm as vector over $\mathbb{Q}$ that is crucial step for AI encoding technique. Therefore combining proposed DCT algorithms and AI encoding technique new error-free computational algorithm of DCT-2 with power of two size can be derived.

The proposed recursive algorithm is also well suited for development of new parallel-pipeline architecture of DCT processor. It can be used by automatic code generation programs that search alternative implementations for the same transform to find the one that is best fitted to the desired platform~\cite{Vanhoof:1993}. 
\section{Summary}
An algebraic technique of derivation of fast DCT algorithms is presented. The technique is based on ASP and Galois theory. The main idea behind the approach is to use the filed of rational numbers $\mathbb{Q}$ for a polynomial algebra associated with DCT. The fast DCT algorithm is derived as a result of stepwise decomposition of the polynomial algebra and requires the extension of the base filed. The extension is determined using Galois theory. The proposed technique is applied to the derivation of a fast $\mathrm{DCT}\text{-}4_{2^k}$ algorithm.

\appendix
\section{Scaled $\mathrm{DCT}\text{-}2_{2^k}$ algorithm} 
$\mathrm{DCT}\text{-}2_n$ is arisen from polynomial algebra 
\begin{equation*}
\mathcal{A}_\mathbb{Q} = \mathbb{Q}[x]/(x-1)U_{n-1}(x),\quad b=(V_0,\ldots,V_{n-1}), 
\end{equation*}
where $U$ is Chebyshev polynomial of the second kind.
Using the factorization 
\begin{equation*}
U_{2n-1}(x) = U_{n-1}(x)\cdot 2T_n(x),
\end{equation*}
the algebra $\mathbb{Q}[x]/(x-1)U_{2n-1}(x)$  can be decomposed as
\begin{eqnarray}
&&\mathbb{Q}[x]/(x-1)U_{2n-1}(x) \notag \\ 
&\rightarrow&\mathbb{Q}[x]/(x-1)U_{n-1}(x)\oplus \mathbb{Q}[x]/2T_{n}(x), \label{eq:dct2_2n_decomp}
\end{eqnarray}
that according to (\ref{eq6})--(\ref{eq8}) leads to the following fast  algorithm~\cite{Pusch8d}
\begin{equation}\label{eq:fast_dct2}
\overline{\mathrm{DCT}\text{-}2}_{2n} = L_n^{2n}(\overline{\mathrm{DCT}\text{-}2}_{n}\oplus \overline{\mathrm{DCT}\text{-}4}_{n})B_{2n},
\end{equation}
where $L_n^{2n}$ is the stride permutation matrix and $B_{2n}$ is change of basis matrix. $B_{2n}$ maps basis $b$ to the concatenation $(c,d)$, where $c=d=(V_0,\ldots V_{n-1})$ are the basis for subalgebras in the right-hand side of~(\ref{eq:dct2_2n_decomp}). The first $n$ columns of $B_{2n}$ are 
\begin{equation*}
B_{2n}=\begin{bmatrix}
I_n	&	*	\\
I_n	&	*
\end{bmatrix},
\end{equation*}
since the elements $V_\ell \in b$ for $0\leq \ell < n$ are already contained in $c$ and $d$. The rest entries are determined by the following expressions
\begin{eqnarray}
V_{n+\ell} \equiv&    V_{n-\ell-1}	 &\mod{(x-1)U_{n}}  \label{eq:new_dct2_basis:a}\\ 
V_{n+\ell} \equiv& 	-V_{n-\ell-1}	 &\mod{2T_{n}},  \label{eq:new_dct2_basis:b}
\end{eqnarray}
which yields
\begin{equation*}
B_{2n}=\begin{bmatrix}
I_n	&	 J_n	\\
I_n	&	-J_n
\end{bmatrix}.
\end{equation*}
(\ref{eq:new_dct2_basis:a})--(\ref{eq:new_dct2_basis:b}) can be induced using the following relation $2T_n = V_n + V_{n-1}$, $(x-1)U_{n-1} = V_n - V_{n-1}$ and $V_n = 2xV_{n-1}-V_{n-2}$. Note that decomposition~(\ref{eq:dct2_2n_decomp}) does not require extension of based field $\mathbb{Q}$. This leads to multiplication-free change of basis matrix $B_{2n}$. 

When the size of $\overline{\mathrm{DCT}\text{-}2}_n$ is power of two (\ref{eq:fast_dct2}) can be applied recursively to obtain fast algorithm. Joint use of factorizations (\ref{eq:fast_dct2}) and (\ref{my_alg}) leads to the scaled fast $\overline{\mathrm{DCT}\text{-}2}_n$ recursive algorithm. For $n=16$ this algorithm requires only 17 multiplication.

\bibliographystyle{siam} 
\bibliography{ref_dct} 
 
\end{document}